\documentstyle[11pt, aaspp4,epsf,rotate]{article}

\lefthead{B\"ottcher \& Liang}
\righthead{Neutron star accretion}

\def\Ktwo{K_2 \left( {1 \over \Theta_e} \right)}
\def\Kthree{K_3 \left( {1 \over \Theta_e} \right)}
\def\dWdt{\left( {dW \over dt} \right)}
\def\nussa{\nu_{\rm SSA}}

\begin{document}

\centerline{Accepted for publication in {\it The Astrophysical Journal},
v. 551 (2001)}

\title{Monte Carlo Simulations of Thermal-Nonthermal Radiation from a Neutron 
Star Magnetospheric Accretion Shell}

\author{Markus B\"ottcher\footnote{Chandra Fellow}\altaffilmark{2}
and Edison P. Liang\altaffilmark{2}}

\altaffiltext{2}{Physics and Astronomy Department, Rice University, MS 108, 
6100 S. Main Street, Houston, TX 77005-1892, USA}

\begin{abstract}
We discuss the space-and-time-dependent Monte Carlo code we have 
developed to simulate the relativistic radiation output from compact 
astrophysical objects, coupled to a Fokker-Planck code to determine 
the self-consistent lepton populations. We have applied this
code to model the emission from a magnetized neutron star accretion 
shell near the Alfv\'en radius, reprocessing the radiation from the 
neutron sar surface. We explore the parameter space defined by the 
accretion rate, stellar surface field and the level of wave turbulence 
in the shell. Our results are relevant to the emission from atoll
sources, soft-X-ray transient X-ray binaries containing weakly
magnetized neutron stars, and to recently suggested models of
accretion-powered emission from anomalous X-ray pulsars.
\end{abstract}

\section{Introduction}

The high energy radiation from compact astrophysical objects is
emitted by relativistic or semi-relativistic thermal and nonthermal 
leptons (electrons and pairs) via synchrotron, bremsstrahlung, and 
Compton processes, plus bound-bound and bound-free transitions of 
high-Z elements.  Since Compton scattering is a dominant radiation 
mechanism in this regime, the most efficient and accurate method 
to model the transport of high energy radiation is the Monte Carlo
(MC) technique.  During the past decade we have developed a versatile 
state-of-the-art space-and-time-dependent MC code to model the radiative 
output of compact objects (see e.g. Liang et al. \markcite{liang00}2000). 
Recently we have added the self-consistent evolution of the leptons
using a Fokker-Planck scheme.  The lepton evolution is then coupled
to the photon transport. Since this is the first time that we
report on results obtained with this code, the first part of the
present paper is devoted to a detailed description of the capabilities
of the code (\S 2) and its verification in comparison with previous
work (\S 3).

In the second part of this paper (\S 4), we apply this coupled MC-FP 
code to the study of the reprocessing of blackbody radiation from a 
magnetized neutron star surface by a magnetized spherical shell at the 
Alfv\'en radius. Recent observations of weak-field neutron star binaries, 
such as low-luminosity, X-ray bursters (atoll sources, e.g., 4U~1608-522: 
\cite{zhang96}, 4U~1705-44: \cite{barret96}), bursting soft X-ray 
transients (e.g., Aql~X-1: \cite{harmon96}), or pulsar binary systems 
(e.g., PSR~B1259-63: \cite{tavani96}) indicate that many of them exhibit 
soft (photon index $\gtrsim 2$) power law tails extending beyond 
$\sim 100$~keV, at least episodically, in addition to the thermal 
component at temperatures of $\sim$~a few keV, which presumably 
originates from the stellar surface. The luminosity of this high-energy
tail appears to be anti-correlated with the soft X-ray luminosity
(\cite{bv94,tl96}).

The origin of this high energy tail is unexplained at present.  
It could be due to thermal Comptonization by a hot coronal plasma, 
or it could be due to nonthermal emission.  Tavani and Liang 
(\markcite{tl96}1996) examined systematically the possible sites 
of nonthermal emissions and concluded that the Alfv\'en surface is 
the most likely candidate since the dissipation of the rotation energy 
of the disk is strongest there, due to magnetic reconnection and 
wave turbulence generation.  Here we first focus on particle 
acceleration by wave turbulence.  

Because of the 1-D nature of our code at this stage, the neutron
star is assumed to emit isotropic blackbody radiation, the 
reprocessing magnetospheric plasma is idealized as part of a 
spherical shell at the Alfv\'en radius, and the magnetic field 
is taken to be nondirectional, so that the radiation output is 
isotropic.  Even though this is not a perfect representation of a 
quasi-dipolar magnetosphere and accretion flow, we believe that, 
except for very special viewing angles, our output spectra should 
be a reasonable approximation to the angle-averaged output of the
reprocessing by a hybrid thermal/nonthermal magnetosphere. We 
assume that the leptons are energized by Coulomb collisions with 
virial ions and accelerated nonthermally by Alfv\'en and whistler 
wave turbulence, and cooled by cyclotron/synchrotron, bremsstrahlung, 
and inverse Comptonization of both internal soft photons and 
blackbody photons from the stellar surface. 

The primary focus of the parameter study presented in \S 4 is the 
application to weakly magnetized neutron stars with surface magnetic
fields of $B_{\rm surf} \lesssim 10^{11}$~G. In particular, we will show
that the anti-correlation of the hardness and luminosity of the hard 
X-ray emission with the soft X-ray luminosity is a natural consequence 
of the energetics of particle acceleration and cooling near the Alfv\'en 
radius. We predict that the nonthermal tails in the hard X-ray spectra
of accreting, weakly magnetized neutron stars may extend up to $\sim
1$~MeV. A solid detection and the measurement of the cutoff energy
of these high-energy tails by the INTEGRAL mission, scheduled for launch 
in 2002, will provide important constraints on accretion-based models
for the hard X-ray emission from accreting neutron stars.

In this context, it is interesting to note that Chatterjee, Hernquist, 
\& Narayan (\markcite{chat00}2000; see also \cite{ms95,wang97,ch00}) 
have recently proposed a similar type of accretion-powered emission
for anomalous X-ray pulsars (AXPs), as an alternative to models based 
on magnetic-field decay (\cite{td96}) or residual thermal energy 
(\cite{hh97}). According to Chatterjee et al. (\markcite{char00}2000)
the X-ray emission from AXPs (which generally consists of a soft,
thermal component with $kT \sim 0.3$ -- 0.4~keV plus a hard X-ray 
tail with photon index $\Gamma \sim 3$ -- 4) is powered by 
accretion of material from the debris of the supernova which had
formed the neutron star, onto a the surface of the neutron star,
which possesses a typical pulsar magnetic field of $B_{\rm surf}
\sim 10^{12}$~G. Therefore, we extend our parameter study to
parameter values relevant to accreting pulsars. However, we point
out that in the case of a magnetic field as high as $B_{\rm surf} 
\sim 10^{12}$~G, the assumed shell geometry and the quasi-isotropy 
of the emission from the neutron star surface may be a gross 
over-simplification. However, although consequently the precise 
parameter values used in this region of the parameter space should 
not be taken at face value, our parameter study might still provide
interesting insight into the dependence of the equilibrium electron
and photon spectra on the various input parameters in the 
high-magnetic-field case.

\section{Physics of photon and lepton evolution}

We use the Monte Carlo (MC) technique (Podznyakov, Sobol, \&
Sunyaev \markcite{poz83}1983; Canfield, Howard, \& Liang
\markcite{canfield87}1987; \cite{liang93}; Hua, 
Kazanas, \& Titarchuk \markcite{hua97}1997) to simulate 
relativistic photon transport. We include the full 
(energy-and-angle-dependent) Klein-Nishina cross section for 
Comptonization, relativistic bremsstrahlung from lepton-ion 
and lepton-lepton scattering (\cite{dermer84}), and cyclo-synchrotron 
processes (\cite{brainerd84}) in magnetic fields.  The MC photon 
transport is fully space-and-time-dependent.  Photons are born 
with a certain ``weight'' (which is basically the total 
energy in photons within a spatial region or emitted from 
a photon source, divided by the number of Monte-Carlo 
particles) which is diminished by absorption and escape, 
until the weight drops below a user-specified limit, at 
which point the photon is "killed". In the simulations
shown later in the paper, we specify this energy weight cutoff
as 1/100 of the initial statistical weight of the photon. The 
final results are insensitive to the choice of this energy weight 
cutoff, as long as it is $\ll 1$. Surviving photons are sampled at 
boundaries to provide time-and-frequency-dependent spectral 
output. In addition to self-emitted photons from the plasma, 
soft photons can be injected at zone boundaries and inside volume 
elements. Currently the code can handle 1-D spherical, cylindrical 
or slab geometries with an arbitrary number of spatial zones.  
However since the photon ray tracing is done with full angle 
informations, the generalization to 2-and-3-D transport is 
straight forward.  The maximum number of photon frequency bins 
is 128.  For more details of this code see Canfield et al. 
(\markcite{canfield87}1987) and B\"ottcher \& Liang 
(\markcite{bl98}1998).  A typical MC run with a million 
particles at a Thomson depth of a few takes 10s of minutes on a 
DEC alpha server.  Since the CPU time usage scales as the square 
of the Thomson depth (~number of scatterings), Thomson thick runs 
can be quite time consuming. We are currently developing a random walk 
scheme for Wien photons trapped in Thomson thick zones, which would 
save large amounts of CPU time without introducing too much error.

The lepton population is computed locally in each spatial zone using the
Fokker-Planck approximation (Dermer, Miller, \& Li \markcite{dermer96}1996;
Li, Kusunose \& Liang \markcite{lkl96a}1996a), taking into 
account coulomb and Moller scattering, stochastic acceleration by Alfv\'en 
and whistler wave turbulence, and radiative cooling (plus pair processes 
if necessary).  In general the lepton population consists of a low energy
thermal population plus a high-energy tail truncated at the highest energies 
by radiative cooling.  The photon and lepton evolutions are coupled to 
each other via a quasi-implicit time scheme in which we use an average
of the photon-electron energy exchange rates between two subsequent time
steps to compute the electron cooling rates, in particular due to Compton 
scattering. The Fokker-Planck equation governing the electron evolution
is solved using a fully implicit scheme. In contrast to several other
codes currently used in the literature (e.g., \cite{stern95}; 
\cite{lkl96a}, \cite{mj00}), we solve the evolution 
of the entire electron population with the FP scheme and do not introduce 
any artificial separation between thermal and non-thermal particles. 
Since the lepton distribution typically evolves much faster than the 
photon distribution, each photon cycle contains many lepton cycles. 
The user, however, can always turn off the nonthermal lepton acceleration 
and assume a strict thermal population whose temperature can be computed 
self-consistently from energy balance alone.

In oder to calculate the emissivities and opacities for thermal 
cyclotron, non-thermal synchrotron, and thermal bremsstrahlung 
emission (and only for this purpose), the electron distribution 
calculated with our Fokker-Planck scheme is decomposed into a 
thermal population plus a non-thermal tail. For the thermal 
bremsstrahlung and non-thermal synchrotron emission and 
absorption, the standard expressions from Rybicki \& 
Lightman (\markcite{rl79}1979) are used. For the thermal cyclotron 
emission, we add explicitly over the first 5 harmonics, beyond which 
we use the asymptotic continuum representation of Mahadevan, Narayan 
\& Yi (\markcite{mny96}1996).

Since in many cases of interest to the current investigation the
moderately to strongly magnetized coronal plasma is optically thick 
at low frequencies, $h \nu \ll 1$~keV, due to synchrotron-(self-)absorption, 
we use the following simplification in order to save CPU time: For 
any given frequency $h \nu \lesssim 1$~keV, we compare the absorption 
length $l_{\rm abs}^{\nu}$ to the Compton mean free path, 
$l_{\rm Compt}^{\nu}$ and the radial extent $\Delta R$ of the 
current zone of the Comptonizing region. If

\begin{equation}
l_{\rm abs}^{\nu} < 0.1 \, \min\left\lbrace l_{\rm Compt}^{\nu}, \,
\Delta R \right\rbrace
\label{abs_length}
\end{equation}
any soft photon produced at this frequency has a very small probability
of escaping the current zone or being up-scattered (to a frequency at
which the absorption length will be different so that the above
criterion has to be re-evaluated after the scattering event) before 
being re-absorbed. Thus, in the frequency ranges where 
Eq. (\ref{abs_length}) is fullfilled, the radiation escaping at 
the boundary of that zone will be approximately
given by the respective section of the thermal blackbody spectrum.
Accordingly, our code neglects the volume emissivities in those
frequency ranges and, instead, produces thermal blackbody photons 
at the zone boundaries.

During each photon time step, the code keeps track of the energy
transferred between photons and electrons due to Compton scattering
and cyclotron/synchrotron and bremsstrahlung emission and absorption.
The respective heating and cooling rates are used to scale the energy
loss/gain coefficients of electrons at a given energy $E_e = \gamma \,
m_e c^2$. For the bremsstrahlung energy loss rate, we use the
approximate scaling law $(d\gamma / dt)_{\rm br} \propto - \gamma^{1.1}$
(B\"ottcher, Pohl, \& Schlickeiser \markcite{bps99}1999). Coulomb 
heating/cooling is included using the energy exchange and dispersion 
rates given in Dermer \& Liang (\markcite{dl89}1989). We assume that 
the ions have a pre-specified temperature $kT_p \gtrsim 1$~MeV, which 
is not significantly affected by any changes of the electron temperature.
For the energy exchange due to M\o ller scattering (elastic 
electron-electron scattering) we use the electron energy exchange
and dispersion coefficients given in Nayakshin \& Melia 
(\markcite{nm98}1998).

In addition to Coulomb interactions and radiative losses, we also
account for stochastic (2$^{\rm nd}$ order Fermi) acceleration due 
to Alfv\'en and Whistler waves. For a given background magnetic field $B$, 
the magnitude and spectrum of hydromagnetic plasma wave turbulences are
determined by the parameters $\delta^2 \equiv (\Delta B / B)^2$, where 
$\Delta B$ is the amplitude of the magnetic-field fluctuations, and $q$, 
the spectral index of the turbulence spectrum. We will generally use $q = 
5/3$, characteristic of Kolmogorov turbulence. The electron acceleration
and energy dispersion rates are calculated using the formalism of
Schlickeiser (\markcite{schlickeiser89}1989). However, we have to
take into account that those plasma waves interacting efficiently
with the low-energy, quasi-thermal part of the electron spectrum
will be efficiently damped and absorbed by the process of Landau
damping (e.g., Schlickeiser, Fichtner \& Kneller \markcite{sfk96}1996).
This leads to a strong truncation of the Kolmogorov wave spectrum
above a critical wave number for which the equivalent absorption
depth through the region occupied by the hot plasma exceeds unity.
We take the effect of Landau damping into account by introducing 
an effective absorption depth $\tau_k$ at wave number $k$, and 
correcting the acceleration rate $\dot\gamma_A^0$ of Schlickeiser 
(\markcite{schlickeiser89}1989) due to the ``optically thin''
plasma wave spectrum by an absorption term:

\begin{equation}
\dot\gamma_A = \dot\gamma_A^0 \, { \left( 1 - e^{- \tau_{k_{\rm res}}}
\right) \over \tau_{k_{\rm res}}}.
\label{gam_dot_A}
\end{equation}
Here $\tau_{k_{\rm res}} = \Gamma_{k_{\rm res}} \, t_{\rm A}$, and
$\Gamma_k$ is the Landau damping rate at wave number $k_{\rm res}$ 
at which Alfv\'en waves are resonating preferentially with electrons 
of energy $\gamma$. $t_A = \Delta R / v_A$ is the Alfv\'en crossing
time of the zone. This modified acceleration term can be renormalized
either to correspond to the pre-specified value of $\delta^2$ for electrons
resonating with waves in the weakly damped part of the Alfv\'en wave
spectrum (i.e. at high electron energies), or by specifying a heat input 
rate into the electron ensemble due to resonant wave/particle interactions.

As an option, our code can account for pair production and 
annihilation and the corresponding photon absorption and emission. 
The pair annihilation rates and annihilation radiation emissivities 
are taken from Svensson (\markcite{svensson82}1982), and the 
$\gamma\gamma$ pair production rate of B\"ottcher \& Schlickeiser 
(\markcite{bs97}1997) is used. In addition, simple Compton reflection
schemes, using the Green's functions for reflection off neutral disk
material of White, Lightman \& Zdziarski (\markcite{white88}1988) and
Lightman \& White (\markcite{lw88}1988), can be used to simulate a 
Compton reflection component either off the inner boundary (corresponding
to a quasi-homogeneous slab geometry) or reflecting part of the radiation
escaping at the outer boundary (e.g., corresponding to a cold outer disk). 

Assuming local isotropy of the electron distribution, we solve the
one-dimensional Fokker-Planck equation

\begin{equation}
{\partial n_e (\gamma, t) \over \partial t} = - {\partial \over
\partial\gamma} \left[ n_e (\gamma, t) \, {d\gamma \over dt} \right]
+ {1 \over 2} \, {\partial^2 \over \partial\gamma^2} \left[
n_e (\gamma, t) \, D(\gamma, t) \right],
\label{fokkerplanck}
\end{equation}
where $D$ is the energy dispersion coefficient. To solve 
Eq. \ref{fokkerplanck}, we use an implicit version of the 
discretization scheme proposed by Nayakshin \& Melia 
(\markcite{nm98}1998). We choose a logarithmic spacing 
in electron kinetic energy, $x_i \equiv \gamma_i - 1$. In 
the following the subscript $i$ refers to electron energy, while 
the superscript $n$ refers to time. We define $f_i^n \equiv n_e 
(\gamma_i, t^n) / n_e$ The discritization is then given by

\begin{equation}
f_i^n = f_i^{n+1} + \Delta t \left( {a_{i+1}^{n+1} f_{i+1}^{n+1} 
- a_{i-1}^{n+1} f_{i-1}^{n+1} \over \Delta x_i + \Delta x_{i-1} } 
\right) - \Delta t \left( { \alpha D_{i+1}^{n+1} f_{i+1}^{n+1} - 
2 D_i^{n+1} f_i^{n+1} + \rho D_{i-1}^{n+1} f_{i-1}^{n+1}
\over \alpha \Delta x_i ( \Delta x_{i-1} + \Delta x_i ) } \right),
\label{discrete}
\end{equation}
where $a_i = (d\gamma / dt) (x_i)$, $D_i = [d(\Delta\gamma)^2 / dt]
(x_i)$, $\alpha = 2 / (1 + q)$, $\rho = 2 q / (1 + q)$, with $q =
x_{i+1}/x_i$. The system of equations is supplemented by the 
boundary conditions specified in Appendix A of Nayakshin 
\& Melia (\markcite{nm98}1998). The system of equations (\ref{discrete}) 
is in tridiagonal form and can be easily solved to find the 
electron distribution at time $t^{n+1}$. In order to 
evaluate the energy exchange and dispersion coefficients
$a_i^{n+1}$ and $D_i^{n+1}$ the code performs an energy balance
calculation to find the average electron energy in the ensemble 
after the current time step, which is then used to calculate the 
energy exchange and dispersion coefficients due to M\o ller scattering.
All other coefficients evolve sufficiently slowly (i.e. on time 
scales much longer than the electron-evolution time scale) so that 
the coefficients evaluated under the conditions at the beginning of 
the current time step can be used.

The implicit scheme of Eq. \ref{discrete} is known to approach the
equilibrium solution for the electron distribution exactly, although
the temporal evolution calculated this way becomes inaccurate on short 
time scales. Since in the problems of interest here, the electron 
distributions typically evolve on timescales much shorter than the 
photon distributions, the electrons always attain a distribution close 
to local equilibrium during each photon time step. Hence, the degree 
of accuracy provided by the implicit scheme is sufficient for our purposes.
As an option, the code can perform the same simulations using an explicit
scheme to solve the Fokker-Planck equation. In all test cases, both schemes
produced virtually identical results, but the implicit scheme executes
a factor of $\sim 10$ -- 100 faster because of the larger individual 
Fokker-Planck time steps allowed in this scheme.

However, we did not find an appropriate method to calculate the 
coefficients for the pair production and annihilation rates at 
time step $n+1$. For this reason, we have to correct the electron 
and positron distribution functions for pair production and 
annihilation in an explicit manner. This becomes obviously 
inaccurate in the case of strongly pair producing model situations. 
Thus, our scheme is a significant improvement over existing schemes 
only for pair deficient situations with pair fractions $f_{\rm pair} 
\equiv n_{e^+} / n_p \lesssim$~a few \%. For strongly pair dominated 
situations, we would have to adjust the individual time steps of our 
simulations so far that such simulations would take a prohibitive
amount of computing time.

In the present paper, we focus on results for equilibrium situations, 
for which we let the code evolve until a stable electron distribution 
and photon output spectrum is reached. The convergence of the 
electron and photon spectra can be greatly accelerated if one starts 
out with an appropriate first guess for the equilibrium electron 
temperature. For this reason, we have developed an analytical 
estimate of the equilibrium electron temperature, which is described 
in Appendix A. This analytical estimate is implemented in our code
and can be used to determine the appropriate initial conditions for
our equilibrium simulations. Detailed tests and applications of the 
time-dependent features of the code will be presented in future 
publications.

\section{Tests of the numerical scheme and comparison with Previous Results}

For verification of our code, we have first compared the individual
energy exchange and diffusion rates, $d\gamma / dt$ and $D (\gamma)$,
with those obtained in earlier work. Our numerical energy exchange
and diffusion rates due to Coulomb scattering are in agreement 
with those of Dermer \& Liang (\markcite{dl89}1998)\footnote{note 
that in their Eq. (A2), the last $\gamma^{\ast}$ has 
to be replaced by $(\gamma^{\ast})^2$}. The numerical
values of the M\o ller scattering energy exchange and diffusion 
coefficients were found in good agreement with those of Nayakshin
\& Melia (\markcite{nm98}1998)\footnote{note that there is a '-' 
sign missing in front of the term $\propto (\gamma - \gamma_1)^2$ 
in their Eq. (35)}.

We have verified that in cases in which radiative cooling and
hydromagnetic acceleration are inefficient, our Fokker-Planck 
scheme correctly produces a thermal electron distribution in 
temperature equilibrium with the protons.

In the analysis of coronal energy and radiation transfer, it is
customary to paramatrize energy input and dissipation rates $\dot u$
[ergs~cm$^{-3}$~s$^{-1}$] by the dimensionless compactness, $\l$, 
which in spherical geometry, is defined by $\l_{\rm sph} \equiv 
4 \pi \sigma_{\rm T} \, R^2 \, \dot u / (3 \, m_e c^3)$, where 
$R$ is the radius of the spherical volume. In slab geometry, 
this becomes $\l_{\rm slab} \equiv \sigma_{\rm T} \, H^2 
\dot u / (m_e c^3)$, where $H$ is the thickness of the 
slab. In our simulations, we specify the heating mechanisms 
to be Coulomb heating through the thermal protons and resonant 
wave-particle interactions. Thus, leaving the shape of the 
electron spectrum general, we calculate the respective energy 
dissipation rates as

\begin{equation}
\dot u_{\rm Coul / A} = m_e c^2 \int\limits_1^{\infty} d\gamma \; 
n_e (\gamma) \left( {d \gamma \over dt} \right)_{\rm Coul / A},
\label{heating}
\end{equation}

where the subscript 'Coul/wp' stands for Coulomb scattering and
wave-particle interaction, respectively.
Obviously, the resulting compactness will depend not only on the
energy density in protons and magnetic turbulence, but also on the
current electron distribution (i.e. temperature, if the distribution
is predominantly thermal), and can thus not be specified a priori
as a free parameter. For comparison with previous work, we need to
find appropriate values of the ion temperature and the wave turbulence
amplitude in order to reproduce the dissipation compactness quoted in 
those papers. We define the temperature corresponding to the electron 
spectrum resulting from our simulations by requiring that the
average electron Lorentz factor, $\langle\gamma\rangle = (1 / n_e)
\int d\gamma \, \gamma \, n_e (\gamma)$, be equal to the average
Lorentz factor of a thermal electron population of the respective
temperature,

\begin{equation}
\langle\gamma\rangle_{\rm th} = {K_3 \left( {1 \over \Theta_e} \right) 
\over K_2 \left( {1 \over \Theta_e} \right)} - \Theta_e.
\label{gamma_average}
\end{equation}

To test our energy balance calculations, we have performed a
series of simulations to reproduce the results of the slab
corona model of Dove et al. (\markcite{dove97}1997), in particular
the inset of their Fig. 2, with which we find good agreement.
For those simulations, we specified ion temperatures of
$80$~MeV~$\lesssim kT_i \lesssim 250$~MeV, a magnetic field 
in equipartition with the ion population, and a negligibly low
level of Alfv\'en wave turbulence to reproduce various values 
of the local heating compactness $\l_c$. As mentioned earlier,
our code is rather inefficient in simulating strongly
pair-producing, high-$l_c$ situations because of the explicit
scheme to solve for pair balance. For this reason, we restrict
the applications of the code in its present version to model
situations with pair fractions $\lesssim$~a few~\%.

Li et al. (\markcite{lkl96a}1996a, \markcite{lkl96b}b)
have developed a code solving simultaneously the Fokker-Planck 
equations for both the electron and the photon distribtions in
a homogeneous medium, including Coulomb interactions, thermal
bremsstrahlung, Compton scattering, resonant wave-particle
interactions, and pair production and annihilation. There are
two major differences between their approach and ours: (1) We
use a Monte-Carlo method to solve the photon transport, and (2)
we solve the Fokker-Planck equation for the entire electron
spectrum, while Li et al. (\markcite{lkl96a}1996a, \markcite{lkl96b}b)
split the electron distribution up into a thermal ``bath'' plus
a non-thermal tail, assuming a priori that electrons of energies 
$\gamma < \gamma_{\rm thr} = 1 + 4 \Theta_e$ attain a thermal 
distribution, and that acceleration due to wave/particle
interactions affects only particles beyond $\gamma_{\rm thr}$.
The latter simplification is justified by the argument that
at low electron energies the thermalization time scale is much
shorter than any other relevant time scale, and that long-wavelength
plasma waves, with which low-energy electrons resonate preferentially,
are strongly damped and in energy elequilibrium with the thermal
pool of electrons. In our simulations, both effects are taken into
account self-consistently without making the a-priori assumption
of the existence or development of a non-thermal population.

Li et al. (\markcite{lkl96b}1996b) present two model calculations 
to explain hard tails observed in the hard X-ray / soft $\gamma$-ray 
spectra of Cyg~X-1 and GRO~J0422+32. A spherical region of radius
$R = 1.2 \times 10^8$~cm is assumed. For the case of Cyg~X-1, they
specify a total heating compactness of $\l = 4.5$, and a non-thermal
heat input into suprathermal particles ($\gamma > \gamma_{\rm thr}$)
of $\l_{\rm st} / \l = 0.15$, wich a turbulence amplitude of $\delta^2
= 0.059$. A soft blackbody radiation component at $kT_{\rm s} = 0.1$~keV
from the outer boundary of the sphere is assumed to provide a soft 
radiation compactness of $\l_{\rm s} / \l = 0.07$. The seed Thomson 
depth of the region is $\tau_{\rm p} = 0.7$. In their simulations, 
Li et al. (\markcite{lkl96b}1996b) find a significant suprathermal 
tail in the resulting electron distribution, which leads to an excess 
hard X-ray / soft $\gamma$-ray emission, consistent with the observed 
one. The temperature of the thermal part of their electron ensemble is
found at $kT_{\rm e} = 139$~keV, and the equilibrium pair fraction is
$f_{\rm pair} \approx 1.8$~\%. For the case of GRO~J0422+32 they specify
$\l = 0$, $\l_{\rm st} / \l = 0.06$, and $\l_{\rm s} / \l = 0.08$.
This resulted in a weaker nonthermal electron tail, an equilibrium
temperature of $kT_{\rm e} = 133$~keV, a wave amplitude of $\delta^2
= 0.065$, and a pair fraction of $f_{\rm pair} \approx 3$~\%.

\begin{figure}
\epsfysize=15cm
\epsffile{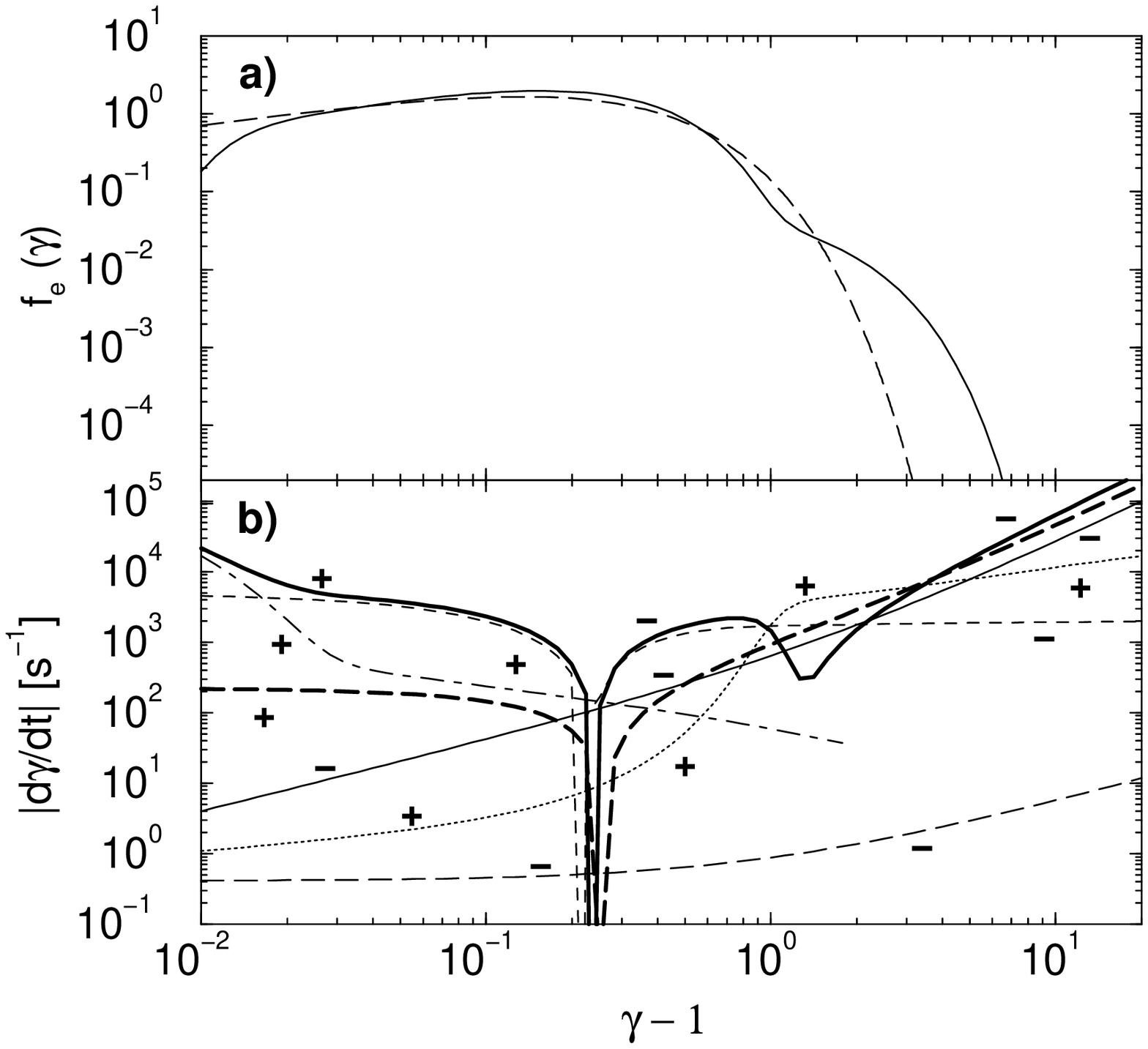}
\medskip
\caption[]{Equilibrium electron spectrum (a) and absolute value of
the energy exchange coefficients (b) for our model calculation with 
parameters similar to the Cyg~X-1 case of Li et al. (\markcite{lkl96b}1996b). 
The labels `+' and `-' in panel (b) indicate the sign of the respective 
contributions. The individual contributions are: synchrotron (thin solid), 
bremsstrahlung (thin long-dashed), Compton (thick long-dashed), Coulomb
scattering (dot-dashed), M\o ller + Bhabha scattering (short-dashed),
hydromagnetic acceleration (dotted), and total (thick solid).
Input parameters were: $kT_i = 3.5$~MeV, $\tau_{\rm p}
= 0.7$, $R = 1.2 \cdot 10^8$~cm, $B = B_{\rm ep} = 1.1*10^6$~G, $\delta^2
= 0.059$, $kT_{\rm BB} = 0.1$~keV, $\l_s = 0.315$. Resulting equilibrium
parameters are $\l = 4.43$, $\l_{\rm st} / \l = 0.46$, $f_{\rm pair} =
1.7$~\%, $kT_e = 105$~keV. The long-dashed curve in panel (a) is a thermal
electron spectrum with $kT_e = 105$~keV.}
\label{fig_cygx1}
\end{figure}

\begin{figure}
\epsfysize=15cm
\epsffile{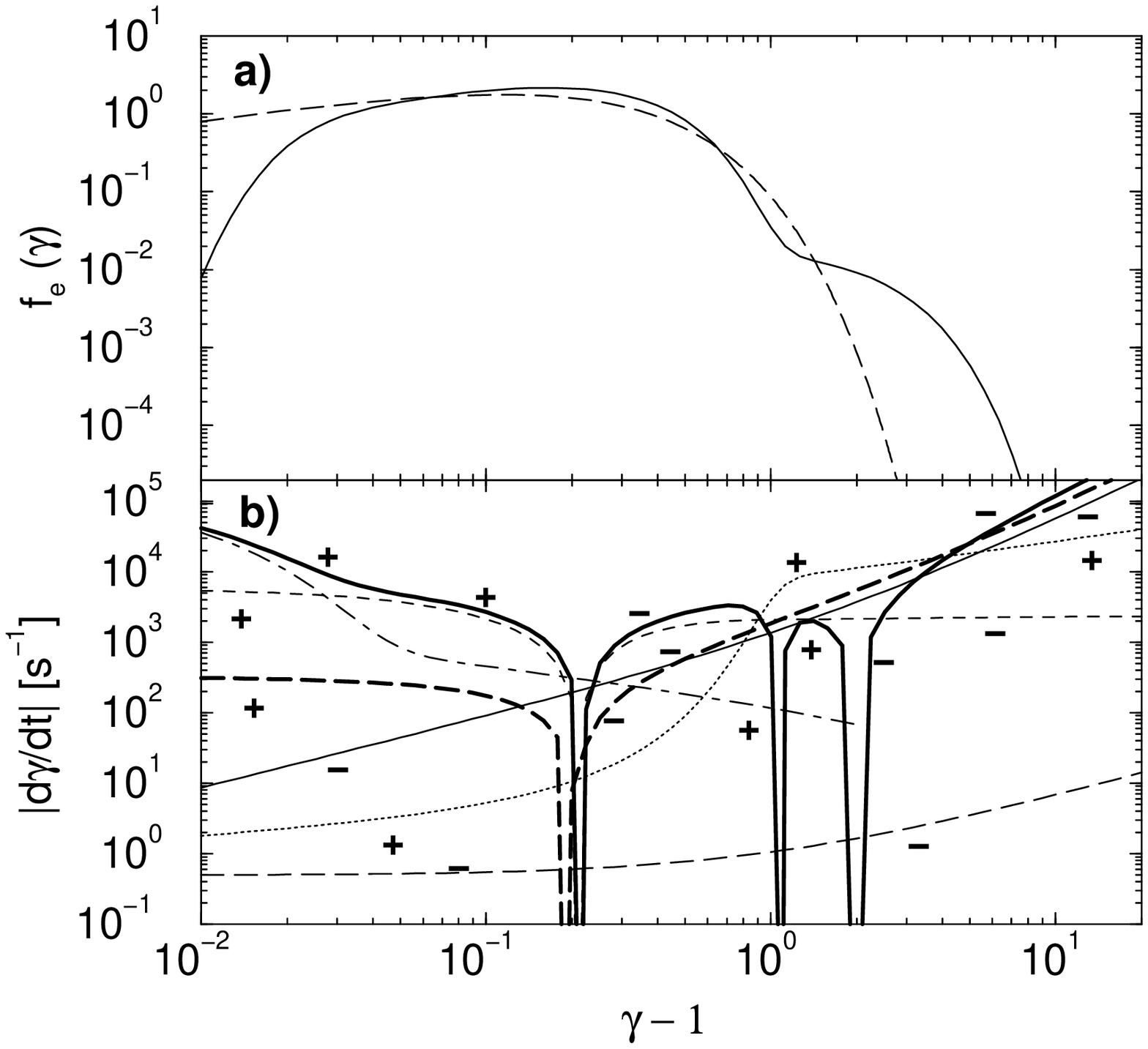}
\medskip
\caption[]{Equilibrium electron spectrum (a) and absolute value of the 
energy exchange coefficients (b) for our model calculation with parameters 
similar to the GRO~J0422+32 case of Li et al. (\markcite{lkl96b}1996b). 
The labels `+' and `-' in panel (b) indicate the sign of the respective 
contributions. The individual contributions are: synchrotron (thin solid), 
bremsstrahlung (thin long-dashed), Compton (thick long-dashed), Coulomb
scattering (dot-dashed), M\o ller + Bhabha scattering (short-dashed),
hydromagnetic acceleration (dotted), and total (thick solid).
Input parameters were: $kT_i = 7.2$~MeV, $\tau_{\rm p}
= 0.7$, $R = 1.2 \cdot 10^8$~cm, $B = B_{\rm ep} = 1.65*10^6$~G, $\delta^2
= 0.065$, $kT_{\rm BB} = 0.1$~keV, $\l_s = 0.72$. Resulting equilibrium
parameters are $\l = 8.66$, $\l_{\rm st} / \l = 0.43$, $f_{\rm pair} =
11$~\%, $kT_e = 93$~keV. The long-dashed curve in panel (a) is a thermal
electron spectrum with $kT_e = 93$~keV.}
\label{fig_0422}
\end{figure}

For comparison with our code, we have run simulations with the same
total compactness $\l$, soft compactness $\l_{\rm s}$, soft blackbody
temperature $kT_{\rm BB}$, radius $R$, seed Thomson depth $\tau_p$, 
and the same values of the plasma wave amplitude normalization $\delta^2$.
A difficulty in comparing the two codes is that in Li et al. 
(\markcite{lkl96a}1996a, \markcite{lkl96b}b) the ion temperature and 
magnetic field are not specified explicitly. In our comparative simulations
we assume a magnetic field in equipartition with the ion energy density.
The results of our simulations are illustrated in Figs. \ref{fig_cygx1}
and \ref{fig_0422}. Our results are qualitatively similar to those of
Li et al. (\markcite{lkl96a}1996a, \markcite{lkl96b}b). However, we find
somewhat lower equilibrium temperatures and stronger nonthermal electron
tails as well as stronger suprathermal acceleration compactnesses 
$\l_{\rm st}$. The lower temperatures may be attributed to a rather 
prominent cyclotron/synchrotron cooling (comparable to Compton cooling) 
in our simulations (see Figs. \ref{fig_cygx1}b and \ref{fig_0422}b). 
This, in combination with the larger $\l_{\rm st}$ values, seems to 
indicate that in our simulations we have used higher magnetic field 
values than Li et al. (\markcite{lkl96a}1996a, \markcite{lkl96b}b).
However, even abandoning the equipartition assumption, we could not
find self-consistent parameter values resulting in the same combination
of input parameters used by Li et al. (\markcite{lkl96a}1996a, 
\markcite{lkl96b}b). Given this descrepancy in the way of input
parameter specification, the qualitative agreement between the
results of the two codes, using very different numerical methods,
is encouraging. 

\section{Models of accretion onto a magnetized neutron star}

While in the first part of this paper we were describing the general
features and the verification of our MC/FP code, we are now applying
this code to model the electron dynamics and photon transport arising
from models of accretion onto a magnetized neutron star. In both weakly
magnetized neutron stars (atoll sources and soft X-ray transient
neutron star binary systems) with $B_{\rm surf} \lesssim 10^{11}$~G
and X-ray pulsars (including, possibly, anomalous X-ray pulsars)
with $B_{\rm surf} \sim 10^{12}$~G, energy dissipation will be 
most efficient at the Alfv\'en radius, where the optically thick,
geometrically thin outer accretion disk is disrupted and the 
dynamics of the accretion flow becomes dominated by the magnetic 
field. We idealize this region of efficient energy dissipation at 
the Alfv\'en radius of disk accretion onto a magnetized neutron star 
as part of a spherical shell whose distance $r_0$, magnetic field, 
and column thickness are fixed by the accretion rate 
(\cite{gl79a},\markcite{gl79b}b,\markcite{gl90}1990).
The distance $r_0$ is given by

\begin{equation}
r_0 = 2 \times 10^8 \, f \, \mu_{30}^{4/7} \, \l_{\ast}^{-2/7} \,
M_{\ast}^{-1/7} \, R_6^{-2/7} \; {\rm cm}
\label{r0}
\end{equation}
where $\mu_{30} = ({\rm neutron \; star \; magnetic \; moment})/(10^{30}
\, {\rm G \, cm}^{3})$, $\l_{\ast} = L / L_{\rm Edd}$, $M_{\ast} = M_{\rm 
NS} / M_{\odot}$, and $R_6 = R_{\rm NS} / (10^6 \, {\rm cm})$. For the
current simulations, for definiteness, we fix the Ghosh-Lamb fudge
parameter $f = 0.3$, and set $M_{\ast} = R_6 = 1$. Hence, the dipole
magnetic field at the Alfv\'en radius is

\begin{equation}
B_0 = 4.2 \times 10^6 \, \l_{\ast}^{6/7} \, \mu_{30}^{-5/7} \, M_{\ast}^{3/7}
\, R_6^{6/7} \, f_{0.3}^{-3} \; {\rm G},
\label{B0}
\end{equation}
where $f_{0.3} = f / 0.3$. The virial ion temperature at $r_0$ is

\begin{equation}
kT_i = {2 \over 3} {G \, M \, m_H \over r_0} \approx 9.3 \, \left(
{r_0 \over 10^7 \, {\rm cm}} \right)^{-1} \, M_{\ast} \; {\rm MeV}.
\label{kTi_virial}
\end{equation}
The column density of the shell can be estimated using the poloidal
accretion rate $\dot M \sim 4 \pi r_0 \, \Delta r_0 \, n_i \, v_p 
\, m_H$, where we assume that the poloidal velocity $v_p \sim 
v_{\rm ff}/2$ with $v_{\rm ff}$ being the free-fall velocity. 
Hence, the radial Thomson depth of the shell is approximately:

\begin{equation}
\tau_T = \Delta r_0 \, n_i \, \sigma_T \sim {\dot M \, \sigma_T
\over 2 \pi r_0 \, v_{\rm ff} \, m_H} \approx 0.97 \, \l_{\ast}^{8/7} 
\, \mu_{30}^{-2/7} \, M_{\ast}^{4/7} \, R_6^{1/7} \, f_{0.3}^{-1/2}.
\label{tauT}
\end{equation}
The neutron star (taken to be a 10 km spherical surface) is assumed to 
emit a blackbody luminosity at the temperature $kT_{\rm BB} = 1.78 \, 
\l_{\ast}^{1/4}$~keV. In addition the level of wave 
turbulence is specified by the dimensionless amplitude $\delta^2 = 
(\Delta B/B)^2$ and the spectral index q.  The minimum wavevector 
$k_{min}$ is set to $2 \pi / (\Delta r_0)$, where $\Delta r_0 \sim
0.1 r_0$ is the shell thickness (\cite{gl79a},\markcite{gl79b}b,
\markcite{gl90}1990).  Due to spherical symmetry of our simulations
the magntic field is assumed to be nondirectional in the shell and 
the synchrotron emissivities and absorption coefficients are 
angle-averaged.

We have explored the parameter space by varying the accretion
rate --- corresponding to a variation of the parameter $\l_{\ast}$
---, the magnetic field --- corresponding to a variation of $\mu_{30}$ 
---, and the amplitude of wave turbulence, $\delta^2$. q is set = 5/3 
in all runs presented in this paper.

\begin{figure}
\epsfysize=12cm
\epsffile{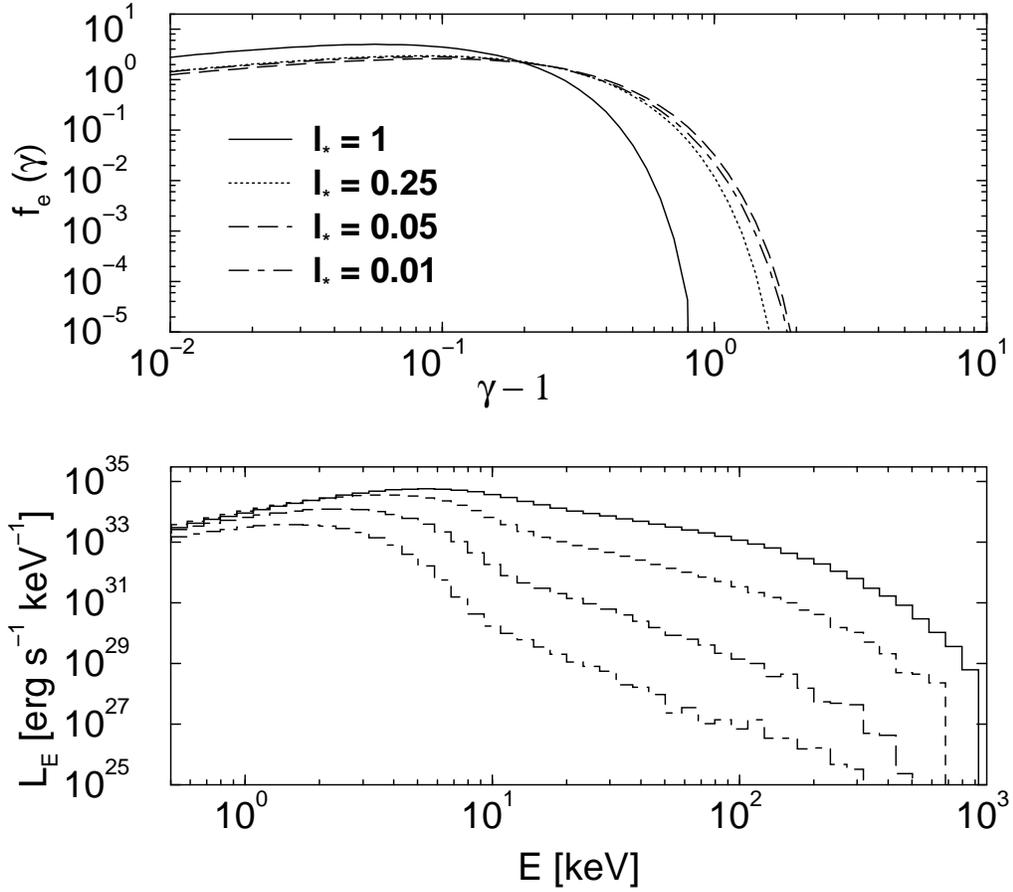}
\medskip
\caption[]{Equilibrium electron spectra (top panel) and photon spectra
(bottom panel) for fixed pulsar dipole moment $\mu_{30} = 1$ and no
plasma wave turbulence ($\delta^2 = 0$).}
\label{fig_delta0}
\end{figure}

\begin{figure}
\epsfysize=12cm
\epsffile{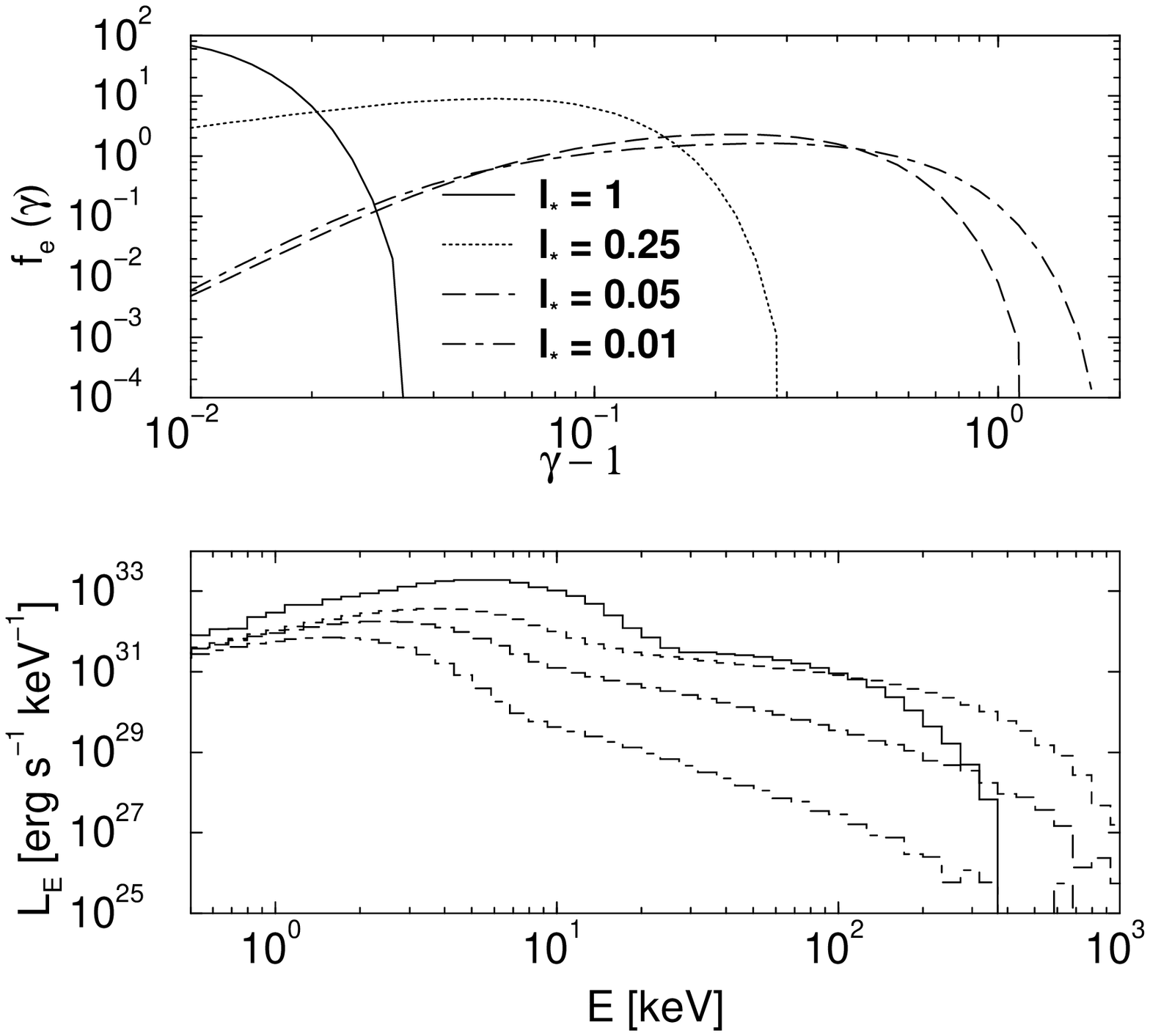}
\medskip
\caption[]{Same es Fig. \ref{fig_delta0}, but with $\mu_{30} = 10^{-3}$.}
\label{fig_delta0_mu_3}
\end{figure}

In Fig. \ref{fig_delta0}, we demonstrate the effect of a varying
accretion rate in the case of no turbulence, $\delta^2 = 0$, and
for a fixed magnetic moment of the neutron star, $\mu_{30} = 1$,
corresponding to a strong surface field of $B_{\rm surf} \sim 
10^{12}$~G. As the accretion rate decreases, the Alfv\'en radius moves
outward, implying a lower magnetic field at the Alfv\'en radius,
and a lower ion temperature and Thomson depth of the active shell. 
At the same time, however, Compton cooling becomes less efficient
due to the reduced soft photon luminosity of the neutron star surface
and due to the larger distance of the active region from the surface.
This reduction of the soft photon compactness leads to an increasing
equilibrium electron temperature in the accretion shell. Consequently,
as the accretion rate decreases, the photon spectra change in the 
following way: For $\l_{\ast} = 1$, the hard X-ray spectrum
smoothly connects to the peak of the thermal blackbody bump from
the neutron star surface, and shows a quasi-exponential cutoff at
high energies. For lower $\l_{\ast}$, the spectrum turns into a
typical low-hard state spectrum of X-ray binaries with the thermal
bump clearly distinguished from a hard power-law + exponential
cutoff at high energies. The hard X-ray power-law becomes harder 
with decreasing $\l_{\ast}$. Fig. \ref{fig_delta0_mu_3} shows the 
same sequence of decreasing $l_{\ast}$ for $\mu_{30} = 10^{-3}$, 
corresponding to $B_{\rm surf} \sim 10^9$~G. At high accretion 
rates, a strong reduction of the equilibrium electron temperature
with respect to the strong-field case results, leading to a softening 
of the photon spectrum with decreasing $\mu_{30}$. This is due to the 
{\it increasing} magnetic field at the Alfv\'en radius as the neutron 
star magnetic moment {\it decreases} (because the Alfv\'en radius 
decreases, overcompensating for the decreasing magnetic moment, see 
Eq. \ref{B0}), resulting in a lower electron temperature caused 
due to increasing cyclotron/synchrotron cooling. For low accretion
rates ($\l_{\ast} \lesssim$~a few \%) the electron and photon spectra
for the strong-field case and the low-field case are virtually
indistinguishable from each other. 

\begin{figure}
\epsfysize=12cm
\epsffile{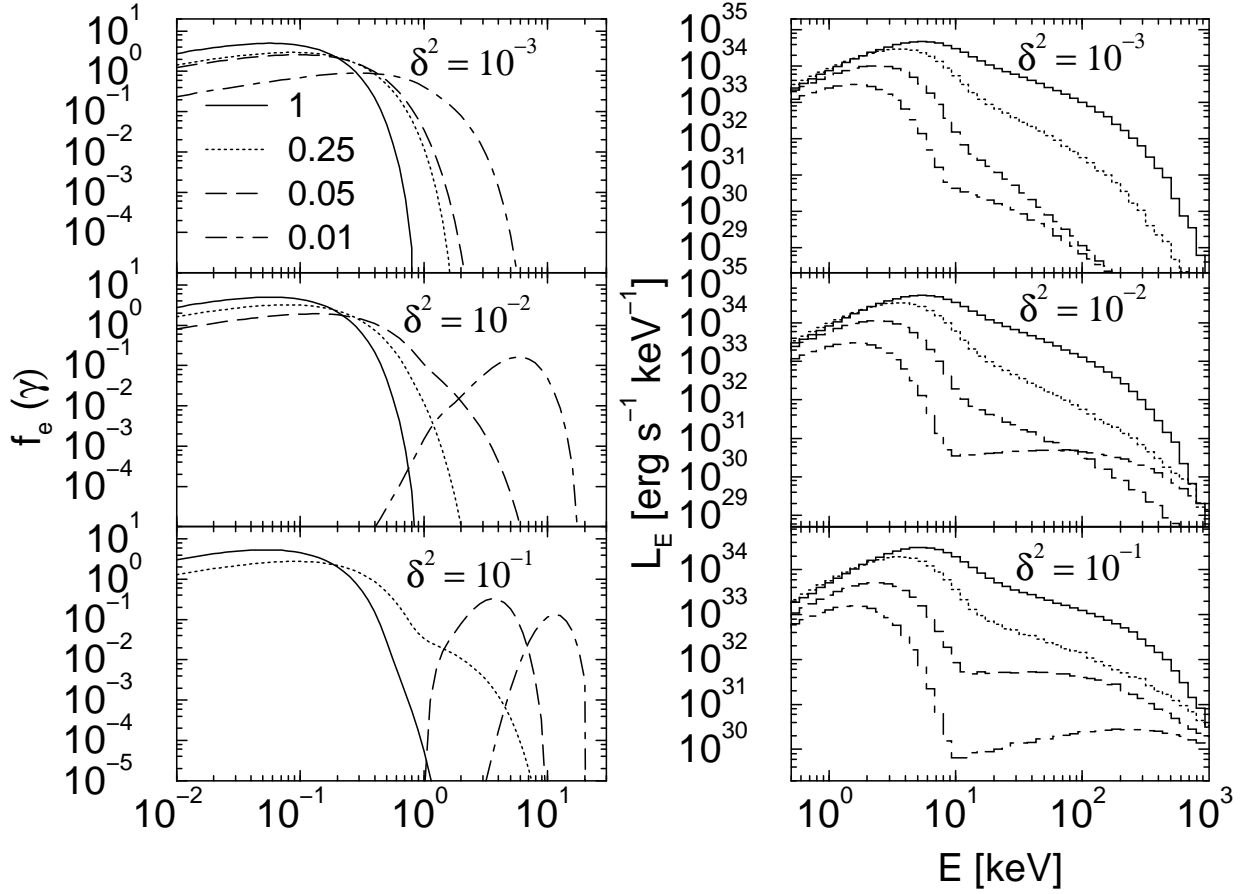}
\medskip
\caption[]{Equilibrium electron spectra (left panels) and photon spectra
(right panels) for fixed pulsar dipole moment $\mu_{30} = 1$. The legend
in the upper left panel refers to the values of $\l_{\ast}$ used in the 
individual runs.}
\label{fig_mu1}
\end{figure}

\begin{figure}
\epsfysize=12cm
\epsffile{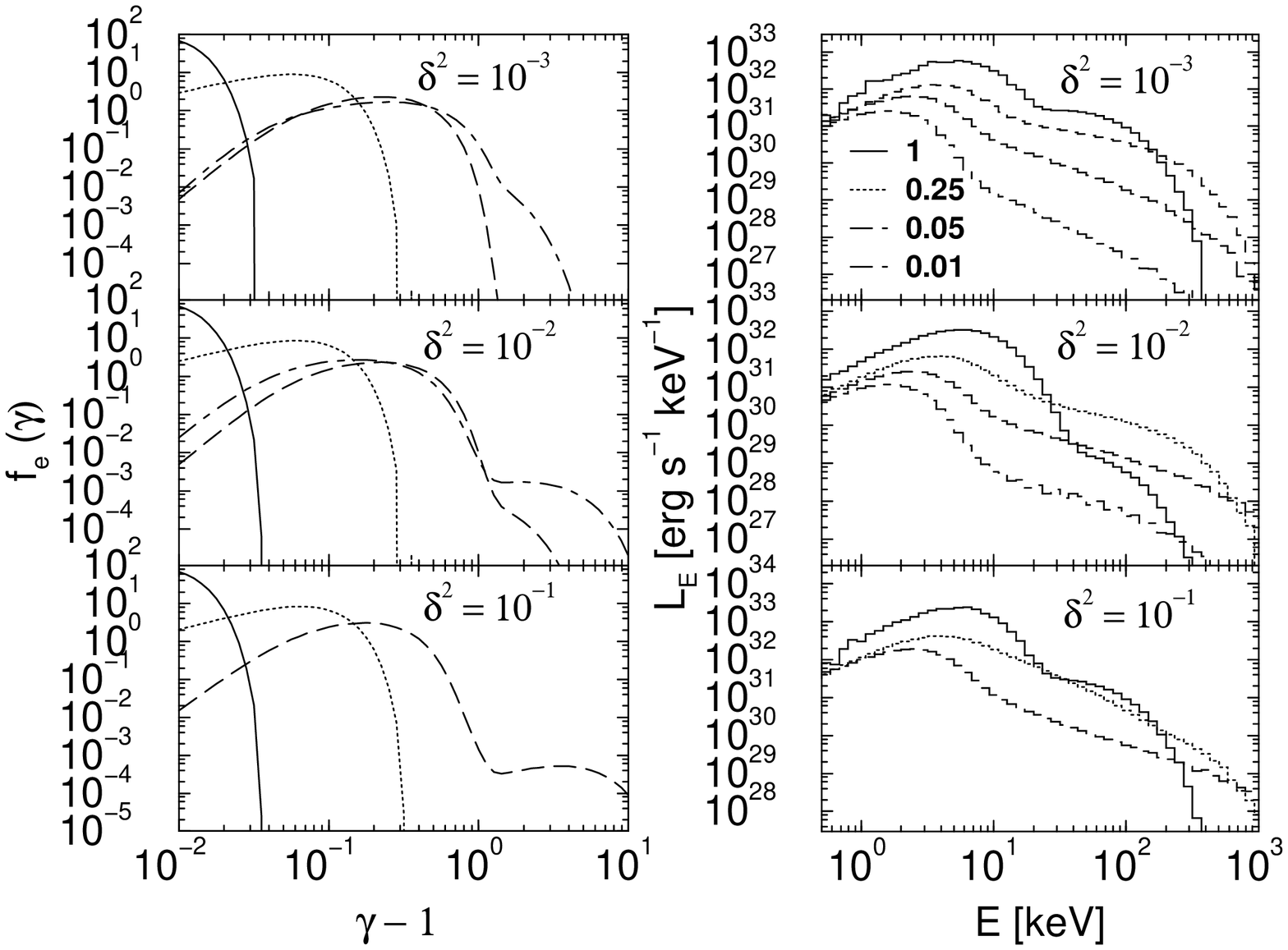}
\medskip
\caption[]{Same es Fig. \ref{fig_mu1}, but with $\mu_{30} = 10^{-3}$.}
\label{fig_mu_3}
\end{figure}

Figs. \ref{fig_mu1} and \ref{fig_mu_3} illustrate the effects
of a varying accretion rate and turbulence level on the electron
and photon spectra. An increasing turbulence level, obviously,
leads to a more pronounced nonthermal tail or bump in the electron
spectrum. At the same time, as a result of increased Compton cooling
on synchrotron photons produced by the nonthermal electrons, the
temperature of the quasi-thermal portion of the electron spectrum
decreases as the turbulence level is increasing. Consequently, the
photon spectrum at soft to medium-energy X-rays becomes softer with
increasing $\delta^2$, while at hard X-ray and $\gamma$-ray energies 
an increasingly hard tail develops. As in the case without turbulence, 
a decreasing accretion rate leads to a higher electron temperature and 
the transition from a smoothly connected thermal blackbody + thermal 
Comptonization power-law spectrum to a typical low/hard state X-ray 
binary spectrum. 

\begin{figure}
\epsfysize=12cm
\epsffile{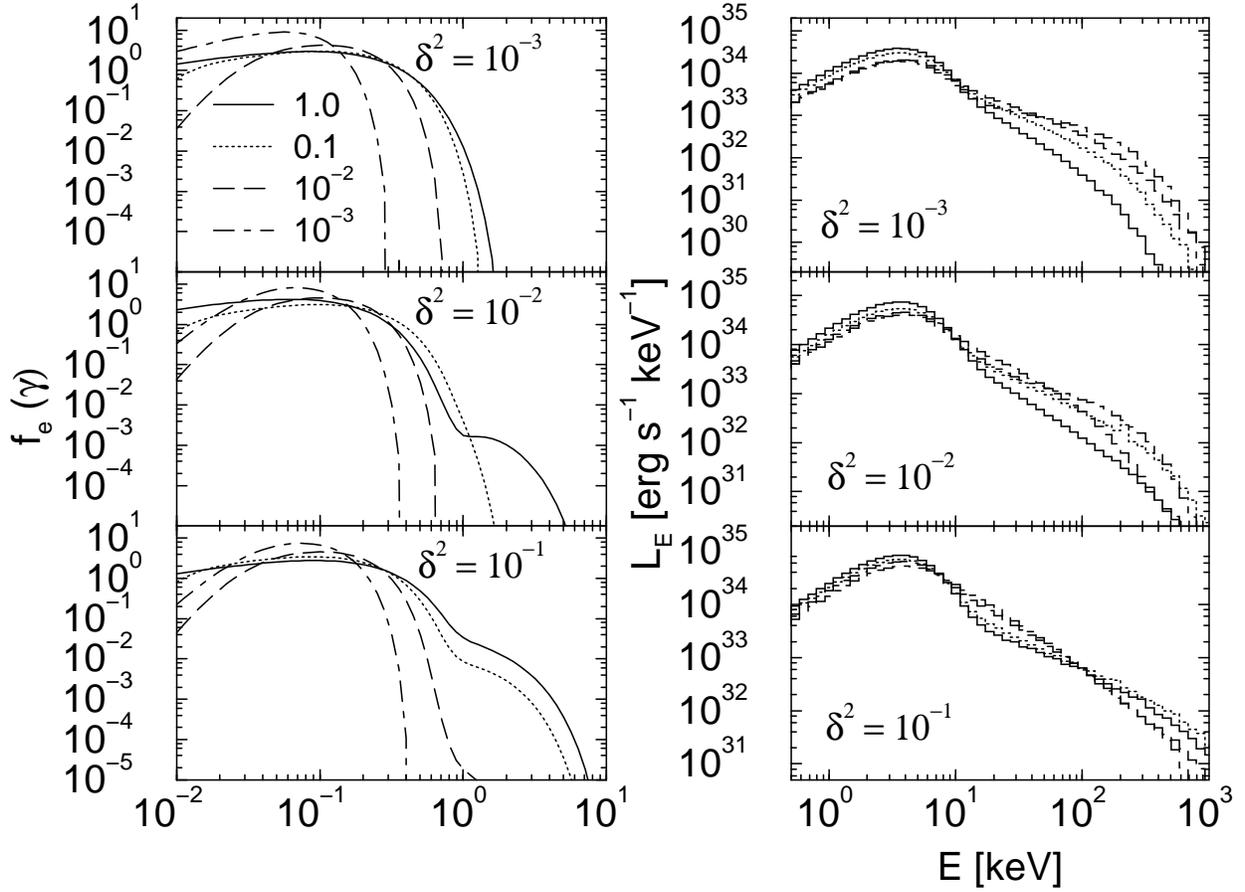}
\medskip
\caption[]{Equilibrium electron spectra (left panels) and photon spectra
(right panels) for fixed accretion rate and luminosity $\l_{\ast} = 0.25$. 
The legend in the upper left panel refers to the values of 
$\mu_{30}$ used in the individual runs.}
\label{fig_l025}
\end{figure}

Fig. \ref{fig_l025} demonstrates the moderate dependence, in particular 
of the resulting photon spectra, on the magnetic moment of the neutron 
star for an intermediate value of the accretion rate, $\l_{\ast} = 0.25$.
Nonthermal tails in the electron spectra become more pronounced with
increasing neutron star magnetic moment. For the weak-field case with
$\mu_{30} = 10^{-3}$, we find that the high-energy end of the electron
spectra are always truncated with respect to a thermal distribution 
due to strong Compton losses, rather than developing a non-thermal tail.

At very low accretion rates, resulting in very low proton and electron
densities in the accretion shell, we find that even at low turbulence 
level the heating due to stochastic acceleration strongly dominates over
Coulomb heating. Consequently, the resulting electron spectrum becomes 
strongly non-thermal. We need to point out that in those cases, our 
treatment is no longer self-consistent since in our code the attenuation 
of Alfv\'en waves is calculated under the assumption that it is 
dominated by Landau damping in a thermal background plasma. 

The results illustrated in Figs. \ref{fig_delta0_mu_3} and
\ref{fig_mu_3} are in excellent agreement with the general trend 
(\cite{bv94,tl96}) that hard tails in LMXBs and soft X-ray transients 
containing weakly magnetized neutron stars are only observed during 
episodes of low soft-X-ray luminosity. While at large soft-X-ray
luminosity, the X-ray emission out to $\gtrsim 20$~keV is dominated
by the thermal component from the neutron star surface and the
hard tail is very soft with a low cut-off energy at $E_{\rm c}
\sim 100$~keV, in lower-luminosity states the hard X-ray tail
becomes very pronounced with photon indices $\Gamma \sim 2$ -- 3
and extends out to $E_{\rm c} \lesssim 1$~MeV. Dedicated deep
observations by the upcoming INTEGRAL mission should be able to
detect this hard X-ray emission from neutron-star-binary soft
X-ray transients and atoll sources in both high and low soft-X-ray
states and thus provide a critical test of this type of accretion
model for weakly magnetized neutron stars.

Interestingly, a similar anti-correlation of the hard X-ray spectral
hardness with the soft X-ray luminosity seems to exist for high surface 
magnetic fields (see Figs. \ref{fig_delta0} and \ref{fig_mu1}) only in 
the case of very low accretion rates ($\l_{\ast} \sim 0.01$) and rather 
high magnetic turbulence levels ($\delta^2 \gtrsim 0.01$). The hard
X-ray spectral indices resulting from our simulations are in excellent
agreement with the values of $\Gamma \sim 3$ -- 4 generally observed
in anomalous X-ray pulsars. This may provide additional support for
accretion-powered emission models for AXPs. Our simulations predict 
cutoff energies of $E_{\rm c} \sim 100$ -- 500~keV.

\section{Summary and Conclusions}

We have reported on the development of a new, time-dependent code for
radiation transport and particle dynamics. The radiation transport,
accounting for Compton scattering, bremsstrahlung emission and absorption,
cyclotron and synchrotron emission and absorption, and pair processes,
is done using a Monte-Carlo method, while the electron dynamics, 
including radiative cooling, Compton heating/cooling, and
stochastic acceleration by resonant interaction with Alfv\'en/whistler
wave turbulence, are calculated using an implicit Fokker-Planck scheme, 
coupled to the Monte-Carlo radiation transfer code. 

In the second part of this paper, we have applied our code to the static 
situation of a shell at the Alfv\'en radius of a magnetized neutron star. 
This situation is representative for accretion onto weakly magnetized
neutron stars (bursting atoll sources or soft X-ray transients containing
wekaly magnetized neutron stars) as well as to recently suggested models
of accretion-powered emission from anomalous X-ray pulsars.

The main results of our parameter study are:

(1) The lepton thermal temperature increases, and the hard X-ray 
photon spectrum becomes harder as the accretion rate is decreasing.
At the same time, the normalization of the hard X-ray power-law,
relative to the thermal blackbody from the neutron star surface,
becomes smaller.

(2) The nonthermal tails in the electron and photon spectra become 
more dominant and harder as the turbulence level is increasing. At
the same time, the quasi-thermal electron temperature decreases,
leading to a softer hard-X-ray spectrum.

(3) For low accretion rates ($\lesssim$~a few \%), the photon spectra 
are only very weakly dependent on the magnetic moment of the neutron 
star. For higher accretion rates, an increasing neutron star magnetic
moment leads to a moderate hardening of the hard X-ray spectrum. 

Our results are in good agreement with the non-detection of hard
X-ray tails during soft X-ray high states of systems containing weakly 
magnetized neutron stars. However, we predict that a hard X-ray excess
beyond $\sim 20$~keV should exist even in the high/outburst state. This
excess should have a cutoff energy of $E_{\rm c} \sim 100$ -- 200~keV.
In the low/quiescent state, the predicted spectral indices of $\Gamma 
\sim 2$ -- 3 are in excellent agreement witht he observed hard X-ray 
excesses observed from sources believed to contain weakly magnetized 
neutron stars. We predict that these spectra should extend out to 
$E_{\rm c} \lesssim 1$~MeV. 

In the strong-field case, representative of a recently suggested model 
for anomalous X-ray pulsars, we expect a strong correlation between
the hard X-ray spectral hardness and the soft X-ray luminosity only
in the case of very low accretion rates and high magnetic turbulence
level. The predicted hard X-ray spectral indices are generally in
very good agreement with the observed values of $\Gamma \sim 3$ -- 4.
Cutoff energies of $E_{\rm c} \sim 100$ -- 500~keV are predicted.

Dedicated, deep observations by the upcoming INTEGRAL mission should
be able to establish the existence of the hard power-law tails predicted
in the models discussed here, and to constrain the high-energy cutoff
of these tails. These measurements will be essential ingredients for a
more detailed modeling of the physical conditions governing the accretion
onto magnetized neutron stars.

\acknowledgements
{The work of MB is supported by NASA through Chandra Postdoctoral 
Fellowship Award No. 9-10007, issued by the Chandra X-ray Center, 
which is operated by the Smithsonian Astrophysical Observatory for 
and on behalf of NASA under contract NAS 8-39073. This work was 
partially supported by NASA grant NAG5-4055. We wish to thank the
referee for very helpful comments, and D. Marsden for pointing out
the relevance of our work to anomalous X-ray pulsars.}

\appendix
\section{\label{appendix_temperature}Estimate of equilibrium temperature}

The code outlined above can be used for equilibrium situations
by simply letting it evolve until both the photon and electron
distributions have relaxed to a steady state. In order to provide
a consistency test of our numerical methods, we compute a
quasi-analytical estimate for the expected equilibrium electron
temperature. In our equilibrium simulations, we use this estimated
equilibrium temperature as initial condition in order to accelerate
the convergence.

Assuming that the electron distribution is roughly thermal,
we may estimate the heating and cooling rates due to the
various processes as follows. Let

\begin{equation}
W = {1 \over n_e} {d E \over d V} = {3 \over 2} k \, T_e
\label{w}
\end{equation}
be the average energy per electrons. We approximate the cyclotron
emissivity by the high-frequency continuum limit given in Mahadevan
et al. (\markcite{mny96}1996):

\begin{equation}
j_{\nu} (\Theta_e) = {2^{1/6} \over 3^{5/6}} \, {\pi^{3/2} e^2 \, n_e
\, \nu \over c \, \Ktwo \, v^{1/6}} \> \exp\left[ - \left( {9 v
\over 2} \right)^{1/3} \right],
\end{equation}
where $\Theta_e = kT_e / (m_e c^2)$, $v \equiv \nu / (\nu_c \,
\Theta_e^2)$,
and $\nu_c = e B / (2 \pi \, m_e c)$.

Denoting by $\Delta R$ the thickness of the emitting shell or slab, the
synchrotron-self-absorption frequency is then determined by solving for

\begin{equation}
1 = \Delta R \, \kappa_{\rm SSA} = \Delta R \, {j_{\nu} (\Theta_e)
\over B_{\nu} (\Theta_e)}.
\end{equation}
The cyclotron/synchrotron cooling rate may then be estimated as

\begin{equation}
\dWdt_{\rm sy} = {1 \over n_e \, V} \left( A_{\rm surface}
\int\limits_0^{\nussa} d\nu \> B_{\nu} (\Theta_e) + V
\int\limits_{\nussa}^{\infty} d\nu \> j_{\nu} (\Theta_e) \right),
\label{Wsy}
\end{equation}
where $V$ and $A_{\rm surface}$ are the volume and the surface area of
the emitting shell or slab.

Assuming that the photon field inside the emitting volume is dominated
by low-energy photons with $\epsilon \ll \Theta_e$, where $\epsilon
= h \nu / (m_e c^2)$, the Compton cooling rate may be approximated as

\begin{equation}
\dWdt_{\rm IC} = - 4 c \, \sigma_T \, u_{\rm ph} \, \Theta_e \,
{\Kthree \over \Ktwo}.
\label{Wic}
\end{equation}
The photon energy density $u_{\rm ph}$ is calculated iteratively
from the contributions of the various soft photon fields, and 
repeated Compton scatterings: Assume that the soft photon energy 
density $u_{\rm s} = u_{\rm sy} + u_{\rm br} + u_{\rm ext}$ 
(sycnhrotron + bremsstrahlung + external radiation field) is 
centered around a soft photon energy $\overline\epsilon_s$, 
and that $\tau_{\rm T} \lesssim 1$. The average energy change 
of a photon with mean photon energy $\overline\epsilon_k$
(Compton scattered $k$ times) changes on average by a factor
$\Delta\overline\epsilon_k = \overline\epsilon_k \, (4 \Theta -
\overline\epsilon_k)$ in the course of the $k+1$. scattering. Then,
the total internal photon energy density is

\begin{equation}
u_{\rm ph} = u_{\rm s} \left( 1 + \sum\limits_{n = 1}^{\infty}
\tau_{\rm T}^n \prod\limits_{k = 0}^{n - 1} \left[ 4 \Theta_e
- \overline\epsilon_k \right] \right),
\label{uph}
\end{equation}
where $\overline\epsilon_0 = \epsilon_s$ and $\overline\epsilon_{k + 1}
= \overline\epsilon_k \, (1 + 4 \Theta_e - \overline\epsilon_k)$.

In the cases we are interested in, the bremsstrahlung photon input
will be negligible compared to cyclotron/synchrotron and the external
soft photon fields. Thus, we assume $u_{\rm s} = u_{\rm ext} + u_{\rm
sy}$
with
\begin{equation}
u_{\rm ext} = {F_{\rm ext} \over c}
\label{u_ext}
\end{equation}
and
\begin{equation}
u_{\rm sy} = {1 \over c} \int\limits_0^{\nussa} d\nu \> B_{\nu}
(\Theta_e)
+ {\Delta R \over c} \int\limits_{\nussa}^{\infty} d\nu \> j_{\nu}
(\Theta_e).
\label{u_sy}
\end{equation}
The photon energy $\overline\epsilon_s$ is $\overline\epsilon_{\rm ext}$
if $u_{\rm ext} > u_{\rm sy}$, or $\max\lbrace\epsilon_{\rm SSA},
\epsilon_0 \rbrace$ else, where $\epsilon_0 = (h \nu_c / m_e c^2)$.

The bremsstrahlung cooling rate is

\begin{equation}
\dWdt_{\rm br} = - \cases{ {128 \over 3 \sqrt{\pi}} \alpha \, r_e^2 \,
m_e c^3 \, n_e \, \sqrt{\Theta_e} & for $\Theta_e \ll 1$ \cr
96 \, \alpha \, r_e^2 \, m_e c^3 \, n_e \, \Theta_e \left( \ln[2
\Theta_e] + 0.673 \right) & for $\Theta_e \gtrsim 1$ \cr}
\label{Wbr}
\end{equation}
(\cite{haug85}) where $\alpha = 1/137$ is the fine structure constant
and $r_e$ is the classical electron radius. 

The heating rate due to Alfv\'en wave turbulence is either specified
as a fixed parameter, or can be evaluated using the acceleration rates
according to Eq. (\ref{gam_dot_A}) as

\begin{equation}
\dWdt_{\rm A} = m_e c^2 \int\limits_1^{\infty} d\gamma \, \dot\gamma_A
\, f_e (\gamma).
\label{WA}
\end{equation}

Finally, the Coulomb heating / cooling rate is
\begin{equation}
\dWdt_{\rm Coul} = {3 \over 2} {m_e \over m_p} \ln\Lambda \, n_{\rm p}
\, c \, \sigma_T \, (k T_{\rm p} - k T_{\rm e}) \, h(\Theta_e,
\Theta_{\rm p}),
\label{WCoul}
\end{equation}
where
\begin{equation}
h(\Theta_e, \Theta_{\rm p}) = { \sqrt{\Theta_e} \over (\Theta_e +
\Theta_{\rm p})^{3/2}} {2 \, (\Theta_e + \Theta_{\rm p})^2 + 2 \,
(\Theta_e + \Theta_{\rm p}) + 1 \over \Ktwo} \, \exp\left( - {1 \over
\Theta_e} \right)
\label{h}
\end{equation}
(\cite{dermer86}), $\ln\Lambda$ is the Coulomb logarithm, $n_p$ is the
number density of ions, and $\Theta_p = kT_p / (m_p c^2)$, .

The total heating/cooling rate consists the sum of all elementary
processes,

\begin{equation}
\dWdt_{\rm total} = \dWdt_{\rm Coul} + \dWdt_{\rm A} + \dWdt_{\rm sy} 
+ \dWdt_{\rm IC} + \dWdt_{\rm br},
\label{Wtotal}
\end{equation}
and we are iteratively solving for the equilibrium solution(s) with
$\dWdt_{\rm total} = 0$.

\newpage

\newpage

\newpage

\newpage

\newpage

\newpage

\newpage

\end{document}